\documentclass[aip,reprint,english,a4paper]{revtex4-1}

\usepackage[pdftex]{graphicx}
\usepackage{amssymb}
\usepackage{amsfonts}
\usepackage{amsmath}

\begin{document}

\title{Real-time observation of ligand-induced allosteric transitions in a PDZ domain}

\author{Olga Bozovic}
\affiliation{Department of Chemistry, University of Zurich, Switzerland}
\author{Claudio Zanobini}
\affiliation{Department of Chemistry, University of Zurich, Switzerland}
\author{Adnan Gulzar}
\affiliation{Department of Chemistry, University of Zurich, Switzerland}
\author{Brankica Jankovic}
\affiliation{Department of Chemistry, University of Zurich, Switzerland}
\author{David Buhrke}
\affiliation{Department of Chemistry, University of Zurich, Switzerland}
\author{Matthias Post}
\affiliation{Institute of Physics, University of Freiburg, Germany}
\author{Steffen Wolf}
\affiliation{Institute of Physics, University of Freiburg, Germany}
\author{Gerhard Stock}
\email{stock@physik.uni-freiburg.de}
\affiliation{Institute of Physics, University of Freiburg, Germany}
\author{Peter Hamm}
\email{peter.hamm@chem.uzh.ch}
\affiliation{Department of Chemistry, University of Zurich, Switzerland}

%

\begin{abstract}
  While allostery is of paramount importance for protein
    regulation, the underlying dynamical process of ligand (un)binding
    at one site, resulting time evolution of the protein structure,
    and change of the binding affinity at a remote site is not well
    understood. Here the ligand-induced conformational transition in a
    widely studied model system of allostery, the PDZ2 domain, is
    investigated by transient infrared spectroscopy accompanied by
    molecular dynamics simulations. To this end, an azobenzene
  derived photoswitch is linked to a peptide ligand in a way that its
  binding affinity to the PDZ2 domain changes upon switching, thus
  initiating an allosteric transition in the PDZ2 domain protein. The
  subsequent response of the protein, covering four decades of time
  ranging from $\sim$1~ns to $\sim$10~$\mu$s, can be rationalized
    by a remodelling of its rugged free energy landscape, with very
  subtle shifts in the populations of a small number of structurally
  well defined states. It is proposed that structurally and
  dynamically driven allostery, often discussed as limiting scenarios
  of allosteric communication, actually go hand-in-hand, allowing the
  protein to adapt its free energy landscape to incoming signals.
\end{abstract}

\date{\today}

\newcommand{\SIFigMass} {S1}
\newcommand{\SIFigITC} {S2}
\newcommand{\SIFigCDFlou} {S3}
\newcommand{\SITableKd} {S1}
\newcommand{\SIFigDecaySpectrum} {S4}

\newcommand{\SIdistances}         {S5}
\newcommand{\SIDBC}     	  {S6}
\newcommand{\SIcontactMap}        {S7}
\newcommand{\SILigandStructure}   {S8}
\newcommand{\SILigandDynamics}    {S9}
\newcommand{\SIMSM}               {S10}
\newcommand{\SIstateDistanceDist} {S11}

\newcommand{\R}[1] {  {\color{red} #1}}
\newcommand{\G}[1] {{\color{magenta} #1}}
\newcommand{\B}[1] {  {\color{blue} #1}}
\newcommand{\F}[2] { \emph{ {\color{blue} \sout{#1} #2} } }

\maketitle


\section*{Introduction}

Allostery represents the coupling of two sites in a protein or a protein
complex, where the binding of a ligand to the distal site modifies the
affinity at the active site.\cite{Wodak2019} Since biological function
is intimately related to protein structure, ligand-induced changes of
the protein's function (e.g., the transition from an inactive to an
active state) are often associated with a change of the protein's mean
structure.\cite{Changeux2012} On the other hand, ligand (un)binding
may also alter the protein's flexibility, which changes the variance
of the structure and gives an entropic contribution to the free
energy.\cite{Cooper84} Referring to the associated change of the
structural fluctuations, the latter scenario, termed ``dynamic
allostery,'' has been invoked to explain apparent absence of
conformational change upon ligand (un)binding. \cite{Cooper84,
  Fuentes06,Bahar07,Petit09,McLeish13,Nussinov15,Guo16,Thirumalai19} Studying the
effects of dynamic allostery has been mainly done by NMR
spectroscopy\cite{Palmer2004,Mittermaier2006,Bourgeois2005}
which, however, only accounts for equilibrium dynamics.

While both models, structural change vs.\ dynamic change, may appear plausible, the nature of the ``allosteric signal'' is not known.
A stringent examination ultimately requires us to study the genesis of
allostery. This includes three steps: (1) The (un)binding of a ligand
(usually initiated by a change of its concentration\cite{Deupi10})
causes (2) the atoms of the protein to undergo a non-equilibrium time
evolution, which (3) eventually leads to a change of the binding
affinity at a remote site of the protein. This so-called ``allosteric
transition'' is a non-equilibrium process and has been observed
directly only rarely, in part because the smallness of the structural
changes makes the transition pathways challenging to observe
experimentally,\cite{Brueschweiler09} and also because of the
time-scale limitations of molecular dynamics (MD)
simulations.\cite{Hyeon05,Pontiggia15,Smith16} In this work, we outline
an approach to study the first two steps, i.e., the ligand-induced
allosteric transition, employing a PDZ2 domain as model system.

\begin{figure}[t]
	\centering
	\includegraphics[width=.48\textwidth]{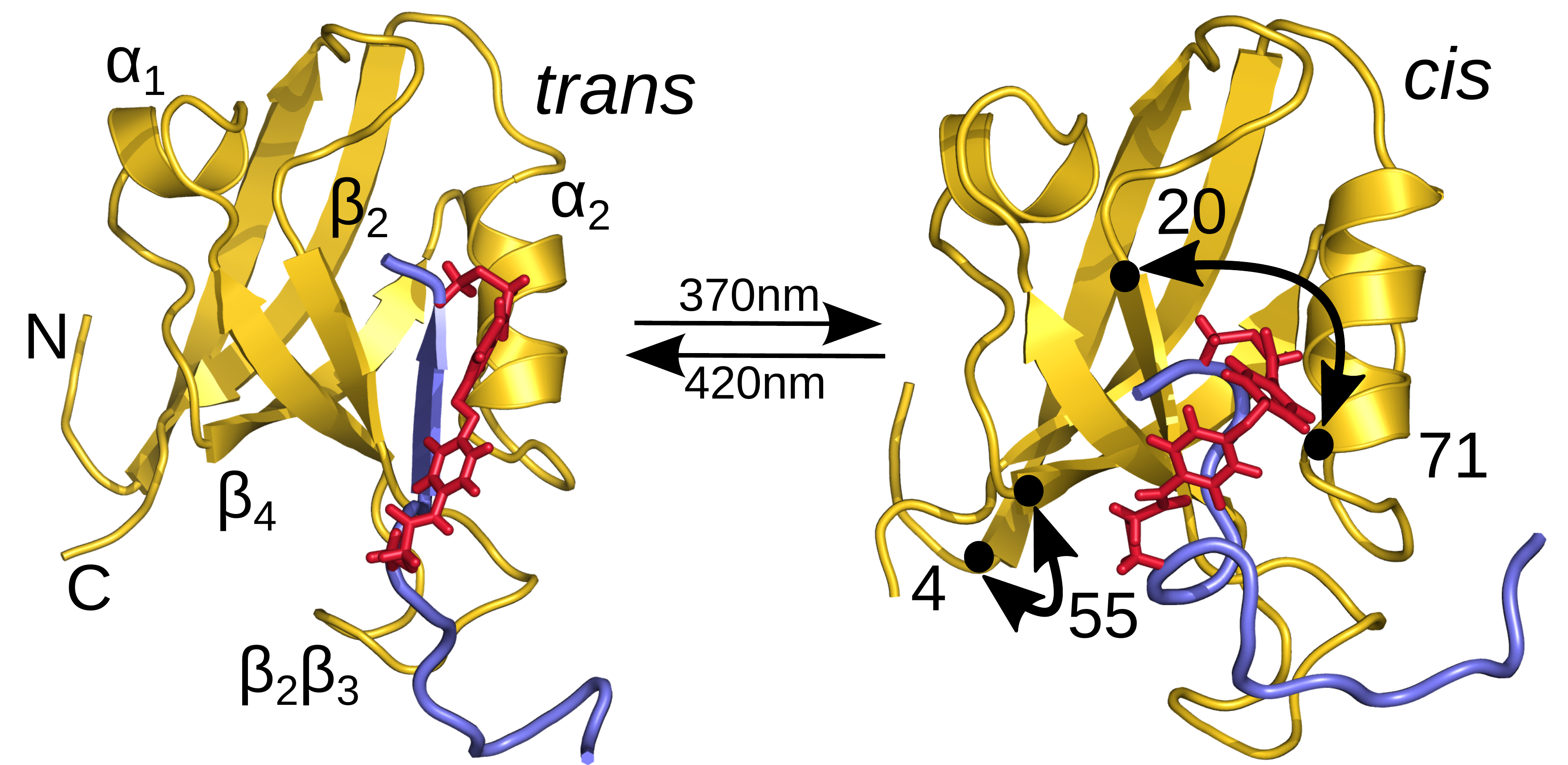}
	\caption{Ligand-switched PDZ2 domain. Main secondary
          structural elements and $C_{\alpha}$-distances $d_{20,71}$
          and $d_{4,55}$ discussed below are indicated. In the {\em
            trans} conformation of the photoswitch (red), the ligand
          (blue) fits well in the binding pocket, while it starts to
          move out when switching to {\em cis}.} \label{FigProtein}
\end{figure}

Known for their modest conformational change upon ligand binding, PDZ
domains are considered as prime examples of dynamic
allostery.\cite{fuentes2004,Fuentes06,Petit09} PDZ
domain-mediated interactions play a pivotal role in many signal
transduction complexes.\cite{Kim2004, Sheng2001} Allosteric
information flow in PDZ domains is thought to be transduced via
conserved allosteric networks in the protein.\cite{lockless99,
 Fuentes06, Law2009, Kong09, reynolds2011}  The system considered here
is the PDZ2
domain from hPTP1E (human tyrosine phosphatase 1E) and a RA-GEF-2
peptide derivative (Ras/Rap1 associating guanidine nucleotide exchange
factor 2)\cite{kozlov02} with an azobenzene moiety linked as photoswitch,\cite{zhang03} see
Fig.~\ref{FigProtein}. It was recently reported for a very
similar system that the phosphorylation of the serine (-2) residue, a
common target in regulatory processes of PDZ domains,\cite{Cao1999}
leads to a \mbox{$\sim$ 5 -- 7-fold} difference in the affinity
towards the PDZ2 domain.\cite{Toto2017} We will see that the binding
affinity can be perturbed to the same extent (\mbox{$\approx$5 fold})
by introducing such a photoswitchable element on the
ligand instead. Since the PDZ2 domain is not modified at all, this strategy
leads to a much less artificial construct than obtained in our previous
study,\cite{Buchli13} where the photoswitch was covalently linked
across the binding pocket of the PDZ2 domain. In addition, using the
ligand as a trigger, one can apply this strategy to virtually any
system.

By photo-isomerizing the azobenzene moiety, we change the binding
affinity of the ligand at a precisely defined point in time. We employ
time-resolved vibrational spectroscopy in connection with a isotope
labeling strategy to monitor the
structural change of the protein in real time, and perform extensive
(more than 0.5 ms aggregate simulation time) all-atom non-equilibrium
MD simulations combined with Markov modeling to interpret the
experimental results in terms of the structural evolution of the
system.  We find that the mean structural change of the protein is
rather small. Yet, in both
experiment and MD simulations the free energy surface of the protein
can be characterized by a small number of metastable conformational
states. In agreement with the view of allostery as an interconversion
between the relative population of metastable
states, we see how
the ligand-induced response of the PDZ2 domain is best described as
remodelling of the free energy landscape,\cite{Frauenfelder91,Dill97,
  smock2009,Tsai2014,Hilser2012} and how the response is
transduced from the ligand to the protein without introducing a
significant structural change.

\begin{figure*}[t]
	\centering
	\includegraphics[width=0.9\textwidth]{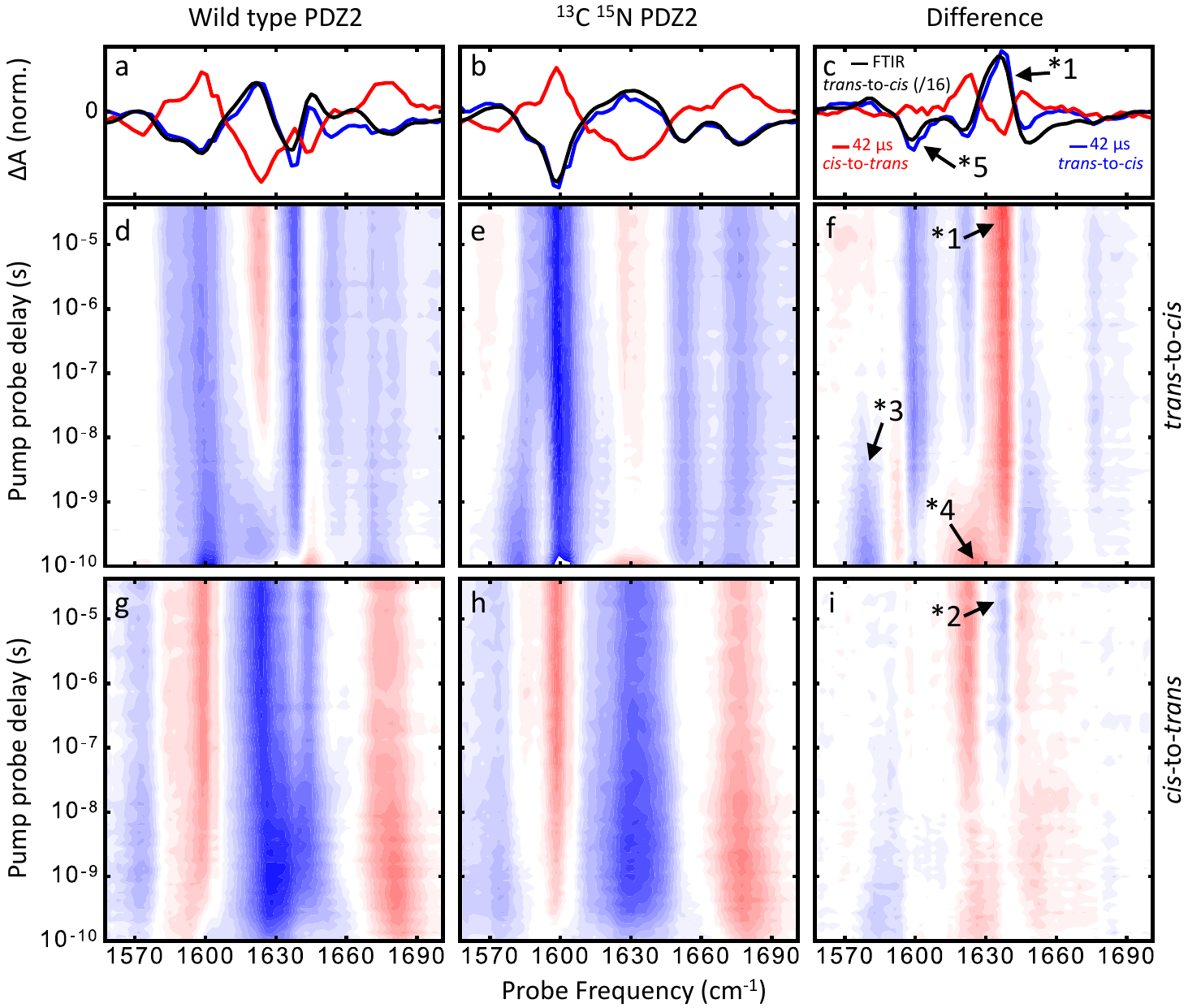}
	\caption{Transient IR spectra of PDZ2 in the region of the
          amide~I band. Panels (a-c) compare transient data at long
          pump-probe delay times (averaged from 20~$\mu$s to 42~$\mu$s
          to increase signal-to-noise) for
          \textit{trans}-to-\textit{cis} (blue) and
          \textit{cis}-to-\textit{trans} (red) switching, together
          with a properly scaled \textit{trans}-minus-\textit{cis}
          FTIR difference spectrum (black). Panels (d-f) show the
          complete transient data for \textit{trans}-to-\textit{cis}
          switching, and panels (g-i) for
          \textit{cis}-to-\textit{trans} switching. Left panels
          show the data for the wild type (WT) protein, middle
          panels for the sample with the protein
          $^{13}$C$^{15}$N labelled (the peptide ligand contains
          naturally abundant $^{12}$C$^{14}$N), and right panels
          the $^{13}$C$^{15}$N-WT difference data. Red colours
          in panels (d-i) indicate positive absorbance changes, blue
          colors negative absorbance changes. The relative scaling of
          the data sets
          and the labelled features are discussed
          in the text.} \label{FigTransientIR}
\end{figure*}

%
%
\section*{Results} 
\subsection*{Experimental}

To set the stage, we have investigated the influence of photoswitching
of the ligand on its binding affinity. By choosing the spacing between
the anchoring points of the azobenzene moiety, the peptide ligand was
designed such that the longer \textit{trans} conformation mimics the
native extended $\beta$-strand conformation, while the \textit{cis}
configuration shortens the peptide and perturbs it from its extended
form. To that end, the alanine residue at position -1 (the ligand is
labelled by negative numbers)
was chosen as the first
anchoring spot for the photoswitch, since it has been shown that a
mutation at this position does not significantly affect the binding,
while residues that are crucial for binding (Val(0), Ser(-2) and
Val(-3)) are preserved.\cite{Chi2012,Lee2010b} The second
anchoring point chosen was Asp(-6) which allows the peptide
to be maximally stretched in the \textit{trans} configuration of the
photoswitch.
Protein and peptide have been expressed/synthesized using standard
procedures,\cite{Zanobini2018,Buchli13} see Materials and Methods for
details. The dissociation constants ($K_D$) in the two
configurations of the photoswitchable peptide were determined by ITC,
fluorescence and CD spectroscopy (see Supplementary
Figs.~\SIFigITC ~and \SIFigCDFlou).\cite{Jankovic2019} The obtained
values averaged for all methods ($K_{D,trans}=2.0\pm0.6~\mu$M,
$K_{D,cis}=9.6\pm0.5~\mu$M, see Supplementary Table~\SITableKd)
reveal an appreciable $\sim\,$5-fold difference in the binding
affinity, with the \textit{cis} state being the destabilized one, as
anticipated.

Considering these binding affinities and the relatively high
concentrations needed for the transient IR experiment (1.25~mM for the
peptide and 1.5~mM for the protein), it is clear that most of the
ligands are bound in both states to a protein of the photoswitch (97\%
in \textit{cis} and 99\% in \textit{trans}), hence we will not observe
many binding or unbinding events.
Furthermore, as binding and unbinding in similar
  PDZ/ligand systems was observed to occur on 10
  -- 100~ms time-scales, \cite{Gianni2005} these processes are hardly within
  the time window of our experiment.
Nevertheless, we will be able to observe the adaptation of the protein to a perturbed peptide conformation in the binding pocket and its transition to unspecific binding on the protein surface.

We investigate that process with the help of transient IR spectroscopy
in the range of the amide I band (see Materials and Methods for
details).\cite{ham00b,Bredenbeck2004,Feng2017} This band originates
from mostly the C=O stretch vibration of the peptide/protein backbone,
and is known to be strongly structure dependent.\cite{barth02} While
one cannot invert the problem and determine the structure of a protein
from the amide I band, any change in protein structure will cause
small but distinct changes in this band (see Fig.~\ref{FigTransientIR} a-c).

Figure~\ref{FigTransientIR} shows the transient IR response in the
spectral region of the amide I vibration after photoswitching in
either the \textit{trans}-to-\textit{cis} (panels d-f) or the
\textit{cis}-to-\textit{trans} direction (panels g-i). To be directly
comparable, the two data sets were scaled in a way that they refer to
the same amount of isomerizing molecules, and not the same amount of
excited molecules.
The scaling took into account the different pump-pulse energies
used in the experiments (see Materials and Methods),
cross sections (23500~cm$^{-1}$M$^{-1}$
for \textit{trans} at 380~nm vs 2000~cm$^{-1}$M$^{-1}$  for
\textit{cis} at 420~nm)\cite{zhang03}, and isomerization
quantum yields (8\% for \textit{trans}-to-\textit{cis} switching and
62\% for \textit{cis}-to-\textit{trans} switching).\cite{borisenko05}

\begin{figure}[t]
\centering
\includegraphics[width=.45\textwidth]{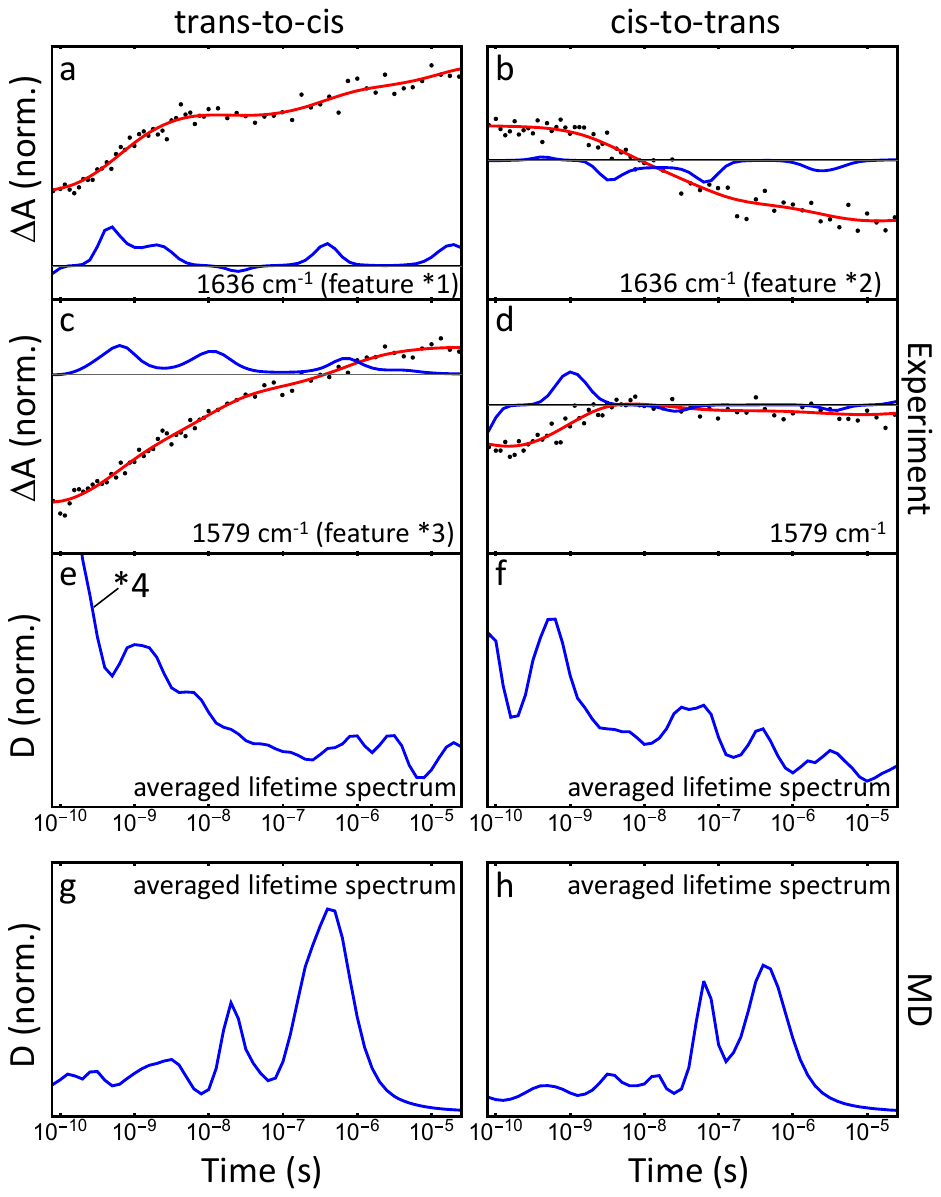}
\caption{Transient $^{13}$C$^{15}$N-WT difference data at
  1636~cm$^{-1}$ (panels a,b) and 1579~cm$^{-1}$ (panels c,d) for
  \textit{trans}-to-\textit{cis} (left) and
  \textit{cis}-to-\textit{trans} (right) switching, highlighting
  features labelled as *1 to *3 in Fig.~\ref{FigTransientIR}.
  Red lines are fits obtained from the time-scale analysis in Eq.\
  (\ref{EqMultiexp}), blue lines represent the resulting
  time-scale spectra $a(\omega_i,\tau_{j})$. Panels (e,f) show the
  corresponding dynamical content; the heat signal labelled as *4 is
  discussed in the text. Panels (g,h) show the MD dynamical content,
  obtained from a time-scale analysis of the non-equilibrium
  time evolution of the mean C$_\alpha$-distances (Supplementary Fig.\
  \SIdistances).} \label{Fig1D}
\end{figure}

The left panels of Fig.~\ref{FigTransientIR} show the results for the
wild type PDZ2 domain, and the middle panels those with the protein
$^{13}$C$^{15}$N labelled, which down-shifts the frequency of the
amide~I band by $\approx$25~cm$^{-1}$. The transient IR responses of
both isotopologues look quite similar, as the signal is dominated by
the photoswitchable peptide, which is perturbed directly by the
azobenzene moiety. To remove that contribution and to isolate the
smaller protein response, the two signals have been subtracted in the
right panels of Fig.~\ref{FigTransientIR}, with some of the more
prominent features highlighted in Fig.~\ref{Fig1D}a-d. In this way, we take advantage of the fact that only the amide~I band of the protein is affected by $^{13}$C$^{15}$N-isotope labelling and not that of the photoswitchable ligand. By doing so, we implicitly assume that the spectra of protein and ligand are additive and that coupling between them can be neglected. Great care was taken that protein and peptide concentrations were exactly the same in both experiments. Furthermore, both experiments were performed right after each other without changing any setting of the laser setup.

Overall, the kinetics of these double-difference spectra are quite
complex and cover many orders of magnitudes in time.\cite{Sabelko99} Furthermore, the
responses for \textit{trans}-to-\textit{cis}
(Figs.~\ref{FigTransientIR}f and \ref{Fig1D}a,c) vs
\textit{cis}-to-\textit{trans} switching (Figs.~\ref{FigTransientIR}i
and Fig.~\ref{Fig1D}b,d) are not mirror-images from each other, which
one might expect if the protein would take the same pathway in the
opposite direction. For example, the strongest band at 1636~cm$^{-1}$
(marked as *1 in Figs.~\ref{FigTransientIR}f and \ref{Fig1D}a) reveals
the biggest step at around 1~ns in the \textit{trans}-to-\textit{cis}
data, while the complementary feature in
\textit{cis}-to-\textit{trans} data (marked as *2 in
Figs.~\ref{FigTransientIR}i and \ref{Fig1D}b) develops in a very
stretched manner from $\approx$3~ns to $\approx$3~$\mu$s. Worthwhile
noting is also a transient band at 1579~cm$^{-1}$ in the
\textit{trans}-to-\textit{cis} data (marked as *3 in in
Figs.~\ref{FigTransientIR}f and \ref{Fig1D}c), living up to
$\approx$100~ns, which has no complementary counterpart in the
\textit{cis}-to-\textit{trans} data (Figs.~\ref{FigTransientIR}i and
\ref{Fig1D}d).

The red lines in Figs.~\ref{Fig1D}a-d are fits revealed from  a
time-scale analysis of the signals using a Maximum Entropy method:\cite{Lorenz-Fonfria2006}
\begin{equation}
  S(\omega_i,t) = a_0(\omega_i)- \sum_k a(\omega_i,\tau_{k})
  e^{-t/\tau_{k}} \label{EqMultiexp}.
\end{equation}
Here $\omega_i$ denotes the probe frequency and $t$ the delay time of
the signal, which is represented by a multiexponential function with
time-scales $\tau_{k}$. The time-scale spectra $a(\omega_i,\tau_{k})$ are shown in Figs.~\ref{Fig1D}a-d
as blue lines.
Each of the kinetic processes discussed above
shows up as a peak in these time-scale spectra, and the pattern of
peaks is different for all the examples shown in
Figs.~\ref{Fig1D}a-d. Nevertheless, the dynamical content,\cite{Stock2018}
$D(\tau_{k}) = [\sum_i a^2(\omega_i,\tau_{k})/n]^{1/2}$,
which averages over the complete data set
shown in Supplementary Fig.~\SIFigDecaySpectrum, seems to indicate a relatively
small number of discrete time scales, see Figs.~\ref{Fig1D}e,f. We
attribute the first peak around 100~ps
(labeled as *4 in
Figs.~\ref{FigTransientIR}f and \ref{Fig1D}e) to a ``heat signal''
originating from the vibrational energy released by the
photo-isomerization of the azobenzene moiety, an effect that is seen
universally in this type of
experiments.\cite{ham97b,Baumann2019}

The transient spectra at the latest pump-probe delay time that is
accessible to our transient experiment (i.e., 42~$\mu$s) are shown in
Figs.~\ref{FigTransientIR}a-c in blue for
\textit{trans}-to-\textit{cis} switching and in red for
\textit{cis}-to-\textit{trans} switching. They are compared to a
properly scaled \textit{trans}-minus-\textit{cis}
          FTIR difference spectrum (black), which represents the response at effectively
infinite time after photoswitching.  The counterpart of the negative
band in the blue and black \textit{trans}-to-\textit{cis} spectra at
1600~cm$^{-1}$ (marked as *5 in Figs.~\ref{FigTransientIR}c) has not
yet evolved in the red \textit{cis}-to-\textit{trans} spectrum. We
conclude from this observation that the \textit{cis}-to-\textit{trans}
transition
is not completely finished after 42~$\mu$s.

%
%
\subsection*{MD simulations}

\begin{figure*}[ht!]
\centering
\includegraphics[width=.99\textwidth]{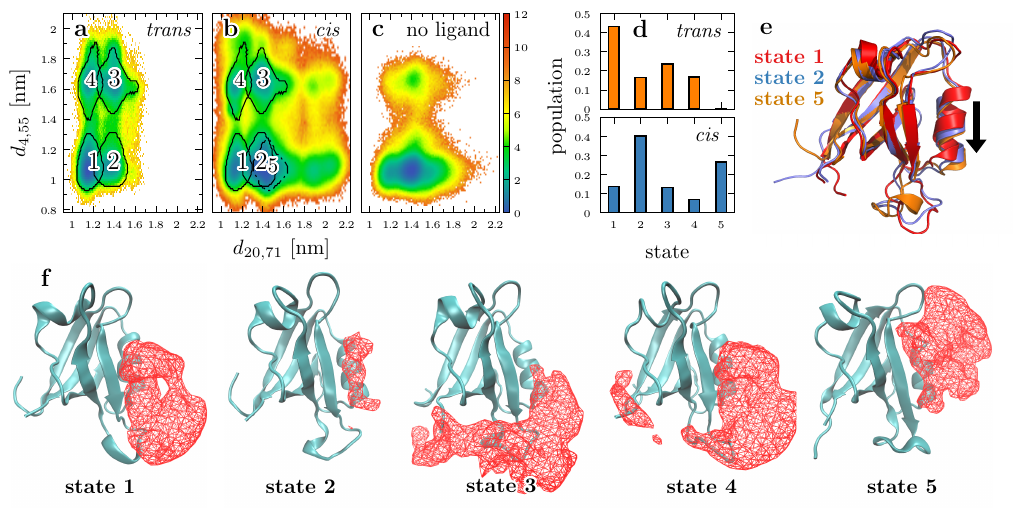}
\caption{
\label{fig:FEL} Identification of metastable
  conformational states. Free energy landscapes (in units of
  $k_{\rm B}T$) obtained from the (a) {\em trans}, (b) {\em cis} and
  (c) ligand-free\cite{Buchenberg14} equilibrium simulations of PDZ2, plotted
  as a function of two essential inter-residue distances. The
  unlabeled state-like feature at the bottom right of (b) represents
  weakly populated ($\lesssim 1\%$) sub-regions of states {\bf 2} and {\bf
    5}. (d) Histogram of the state populations in {\em trans} and {\em
    cis} equilibrium, revealing the ligand-induced population shift of
  PDZ2. (e) Comparison of minimum-energy structures the of states \textbf{1}, \textbf{2} and \textbf{5}, revealing an increased opening of the ligand binding pocket by a downward motion of $\alpha_2$. (f) Structures of states together with
    position densities of the ligand. The isosurface encloses a volume
    with a minimal probability of 0.4 to find a ligand atom within in
    all simulation snapshots belonging to a specific state. Fixed
    points for the comparison are the C$_\alpha$ atoms of strands
    $\beta_4$ and $\beta_6$.}
\end{figure*}

To aid the interpretation of the above experiments, we performed
all-atom explicit-solvent MD simulations of the {\em cis} and {\em
  trans} equilibrium states as well as non-equilibrium MD simulations
\cite{Nguyen06} of the ligand-induced conformational changes of PDZ2.
Using the GROMACS v2016 software package\cite{Abraham15} and the
Amber99*ILDN force field,\cite{Hornak2006, Best2009,
  LindorffLarsen2010} we collected in total $510\, \mu$s simulation
time (see Materials and Methods). For the structural characterization
of the protein, we determined 56 $C_{\alpha}$-distances $d_{i,j}$
between residues $i$ and $j$ that are not redundant (such as $d_{i,j}$
and $d_{i,j\pm 1}$) and whose ensemble average changes significantly
($\langle \Delta d_{ij} \rangle \ge 0.5\,$\AA) during the
non-equilibrium simulations (Supplementary Fig.\ \SIdistances). To
identify the essential coordinates of the system, we performed a
principal component analysis on the normalized distances of all
simulation data,\cite{Sittel2018} followed by robust density-based
clustering\cite{Sittel16} and a recently proposed machine learning
approach\cite{Brandt2018} (see Materials and Methods and Supplementary
Fig.\ \SIDBC\
for details). While we used six dimensions for the clustering, we find
that two $C_{\alpha}$-distances suffice to qualitatively characterize
the conformational distribution of PDZ2: $d_{20,71}$ accounting for
the width of the binding pocket located between $\beta_2$ and
$\alpha_2$, as well as $d_{4,55}$ representing the distance between
N-terminus and $\alpha_1$-$\beta_4$ loop, which reflects the
compactness of the C- and N-terminus region (see Fig.\
\ref{FigProtein}).
Employing these coordinates, Fig.\ \ref{fig:FEL}a shows the free
energy surface $\Delta G = - k_{\rm B}T \ln P(d_{20,71},d_{4,55})$,
obtained from $5 \!\times\! 5 \mu$s-long {\em trans} equilibrium
simulations describing the ligand-bound state of PDZ2. The free energy
landscape reveals four well-defined local minima indicating metastable
conformational states of the system. Density-based clustering
identifies state {\bf 1} as close to the crystal
structure,\cite{Zhang10} while state {\bf 2} indicates an opening of
the binding pocket. Both states are mirrored by states {\bf 3} and
{\bf 4}, which are shifted to larger values of coordinate $d_{4,55}$.

Upon switching the ligand from {\em trans} to {\em cis} configuration,
PDZ2 undergoes a non-equilibrium time evolution until it relaxes within
a few microseconds (see below) into its {\em cis} equilibrium state,
describing the perturbed
protein-ligand complex. Performing
$25 \!\times\! 10\, \mu$s-long {\em trans}-to-{\em cis} non-equilibrium
simulations, we took the last $7 \mu$s of each trajectory to estimate
the rather heterogeneous conformational distribution of the {\em cis}
equilibrium state. When we compare the resulting free energy
landscapes of {\em cis} and {\em trans}, Figs.\ \ref{fig:FEL}a,b
reveal that the accessible conformational space in {\em cis} is
considerably increased, along with the occurrence of additional state
{\bf 5} that reports on a further opening of the binding pocket.
Representing the populations of all states in {\em trans} and {\em
  cis} as a histogram, Fig.\ \ref{fig:FEL}d demonstrates that the
photoswitching of the ligand causes a notable ($\lesssim\,20\,\%$)
shift of the state populations, mostly from state {\bf 1} to states
{\bf 2} and {\bf 5}.

To illustrate the conformational changes associated with these states,
Fig.\ \ref{fig:FEL}e displays an overlay of minimum-energy structures
of states {\bf 1} and {\bf 2} as well as the {\em cis}-specific state
{\bf 5}. We find that the opening of the binding pocket described by
$d_{20,71}$ mainly reflects a shift of the $\alpha_2$ helix down and
away from the protein core. Interestingly, the structural
rearrangement between main states {\bf 1} and {\bf 2} results in an
overall root mean squared (RMS) displacement of only $\lesssim\,1\,$\AA\ and causes only few
($\sim 5$) contacts to change (Supplementary Fig.\ \SIcontactMap). This is in
striking contrast to the cross-linked photoswitchable PDZ2 studied by
Buchli et al.\cite{Buchli13} where 34 contact changes were found for
the {\em trans}-to-{\em cis} reaction,\cite{Buchenberg14} and {\em
  cis} and {\em trans} free energy landscapes hardly
overlapped.\cite{Stock2018} This findings indicate that
ligand-switching is considerably less invasive than a cross-linked
photoswitch and therefore better mimics the natural unbiased system.

Is the above discussed population shift as well as the very occurrence
of states an inherent property of the protein's rugged free energy
landscape,\cite{Frauenfelder91,Dill97} or are these features rather induced
by the ligand? Figure \ref{fig:FEL}c addresses this question
by showing the free energy landscape
obtained from previously performed $6\!\times\!1\mu$s-long simulations
of PDZ2 {\em without} a ligand \cite{Buchenberg14}.  While the state
separation along coordinate $d_{4,55}$ still exists, we find that
states {\bf 1}, {\bf 2} and {\bf 5} merge into a single energy
minimum. It is centered at the position of state {\bf 2}, but is wide
enough to cover a large part of states {\bf 1} and {\bf 5}. Similarly
states {\bf 3} and {\bf 4} form a weakly populated ($2\,\%$)
single minimum. This indicates that ligand-free PDZ2 provides the
flexibility to assess the entire free energy landscape explored during
binding and unbinding, while the interaction with the ligand appears
to stabilize conformational states {\bf 1} and {\bf 4}.

Showing protein structures of the main states together with
position densities of the ligand, Fig.\ \ref{fig:FEL}f
illustrate these interactions (see
also Supplementary Fig.\ \SILigandStructure). For one, we notice that the opening and
closing of the binding pocket (described by $d_{20,71}$) is associated
with the conventional binding of the ligand's C-terminus in this
pocket, which stabilizes closed state {\bf 1} in {\em trans}. In the
open state {\bf 2}, the probability to find the ligand in its
binding mode is significantly decreased, pointing to a
reduced ligand affinity of the protein. On the other hand, we find
that the distinct conformations of the protein's termini described by
$d_{4,55}$ are a consequence of the formation of contacts with the
ligand's N-terminus in states {\bf 3} and {\bf 4}, which are absent in
states {\bf 1}, {\bf 2} and {\bf 5}.  In particular, state {\bf 5}
represents a situation where the hydrophobic photoswitch of the ligand
forms a contact with a hydrophobic bulge at the protein surface around
Ile20, which can be classified as unspecific binding of the ligand to
the protein surface.

\begin{figure}[ht!]
\centering
\includegraphics[width=.49\textwidth]{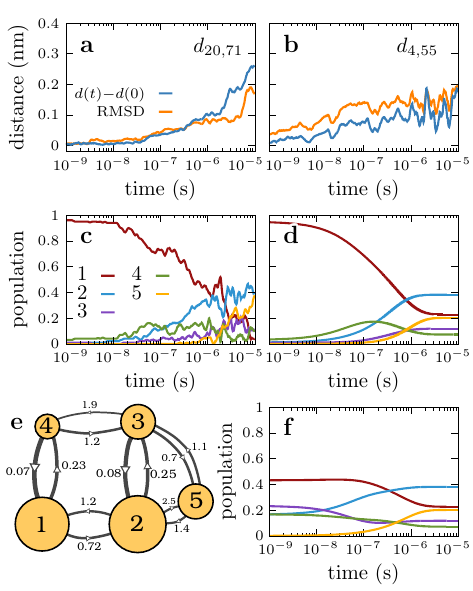}
\caption{
\label{fig:MSM} Time evolution of various
  structural descriptors, following {\em trans}-to-{\em cis}
  ligand-switching of PDZ2. Shown are means (blue) and RMSD (orange)
  of C$_\alpha$-distances (a) $d_{20,71}$ and (b) $d_{4,55}$, as well
  as (c,d,f) populations of conformational states. For easier
  representation, all MD data were smoothed.
  Starting at time $t\!=\!0$ almost completely in state {\bf 1}, we
  compare results from (c) the non-equilibrium MD simulations to (d)
  the corresponding predictions of a Markov state model (MSM). (e)
  Network representation of the MSM. The size of the states indicate
  their population, the thickness of the arrows and numbers indicate
  the transition times (in $\mu$s). For clarity, we discard
  transitions that take longer than $2.5\,\mu$s. (f) MSM simulations
  of the {\em trans}-to-{\em cis} transition, using {\em trans}
  equilibrium initial conditions.}
\end{figure}

Adopting our {\em trans}-to-{\em cis} non-equilibrium simulations, we
can describe the overall structural evolution of PDZ2 in terms of
time-dependent expectation values of various observables. As an
example, Figs.\ \ref{fig:MSM}a,b show the time evolution of the two
C$_\alpha$-distances $d_{20,71}$ and $d_{4,55}$ introduced
above. Following {\em trans}-to-{\em cis} ligand switching, it takes
about 100 ns until the sub-picosecond photoisomerization of the
photoswitch affects the protein's binding region (indicated by
$d_{20,71}$), which becomes wider as the ligand moves out.
The flexible N-terminal region indicated by $d_{4,55}$, on the other hand,
undergoes conformational changes already within a few nanoseconds.
The
  weak correlation between the two inter-residue distances (i.e.,
  $\langle d_{20,71} d_{4,55} \rangle (\langle d_{20,71}^2
  \rangle\langle d_{4,55}^2 \rangle)^{-1/2} \lesssim 0.02$
  for all data), however, indicates that this early motion of the
  terminal region may be not directly related to the functional dynamics
  of PDZ2.  Interestingly, the associated root mean squared
deviations (RMSD) of the two distances show quite similar behavior.
Moreover, Supplementary Fig.\ \SILigandDynamics\ displays various ligand-protein
distances and contact changes, which illustrate that the ligand leaves
the binding pocket on time-scales of 0.1 -- $1\,\mu$s.  When we
calculate the dynamical content of all considered intraprotein
C$_\alpha$-distances, we obtain a time-scale distribution that roughly
resembles the experimental result (Fig.\ \ref{Fig1D}g,h).

%
%

It is instructive to consider the resulting time-dependent populations
of the protein's metastable states. Choosing initial
conditions close to the crystal structure,\cite{Zhang10} Fig.\
\ref{fig:MSM}c exhibits the {\em trans}-to-{\em cis} time evolution of
the state populations. The system starts at time $t=0$ almost
completely in state {\bf 1} and converts to the other
states within microseconds.
To rationalize these findings, we construct a Markov state
model\cite{Bowman2013,Sengupta18} (MSM) which describes the
conformational dynamics of PDZ2 via memory-less jumps between
metastable states. To this end, we calculate a transition matrix $T$
containing the probabilities $T_{ij}$, that the system jumps from
state $i$ to $j$ within lag time $\tau_\text{lag}$, and determine its
eigenvectors $\psi_k$ and eigenvalues $\lambda_k$ (see
Materials and Methods and Supplementary Fig.\ \SIMSM\ for technical
details). As a first
impression, Figs.\ \ref{fig:MSM}c,d compares the state populations
obtained from the non-equilibrium MD simulations and the corresponding
MSM predictions (using $\tau_\text{lag}\!=\!1\,$ns).  We find
excellent agreement for the first three decades of time, but only
qualitative agreement in the last decade, which reflects the bias of
our non-equilibrium MD simulations towards shorter time-scales
($75 \!\times\! 1\,\mu$s-long and $25 \!\times\! 10\,\mu$s-long data).
Showing a network representation of the MSM, Fig.\ \ref{fig:MSM}e
illustrates the connectivity and transition times of the system. We
see that the open-close transition of the binding pocket occurs on a
time-scale of $\sim\,1\,\mu$s, whereas transitions from states {\bf 1}
and {\bf 2} to states {\bf 4} and {\bf 3} are a factor 4 faster with a
back-rate that is even a factor 10 faster.

Assuming a time-scale separation between fast intrastate fluctuations
and rarely occurring interstate transitions, MSM
theory\cite{Bowman2013} states that the time-dependent expectation
value of any dynamical observable can be written as a sum over
exponential functions ${\rm e}^{-t/t_k}$ weighted by the projection of
the observable onto the $k$th eigenvector of transition matrix
$T$. The implied time-scales $t_k \!=\! -\tau_\text{lag}/\ln \lambda_k$
of the MSM therefore govern the time evolution of such different
observables as vibrational spectra and state populations.\cite{Noe2011}
To facilitate a comparison of experimental and simulated time
evolutions, we run a MSM simulation using {\em trans} equilibrium
initial conditions, which is also the starting point of the {\em
  trans}-to-{\em cis} experiments. Comparing the simulation results
(Fig.\ \ref{fig:MSM}f) to the experimental time traces (Fig.\
\ref{Fig1D}), we find that both spectral and population
evolutions appear to be completed on microsecond time-scale.  Moreover,
the MSM populations exhibits various transient features on time-scales
of 10 -- 100 ns, which are also present in the experimental time
signals.

%
%
\section*{Discussion and Conclusions}

Combining transient IR spectroscopy and non-equilibrium MD
simulations, we have described the ligand-induced conformational
transition in the PDZ2 domain, which is thought to be responsible for
protein allosteric communication. We have found that the free energy
landscape of PDZ2 can be described in terms of a few metastable
states with well-defined structure (Fig.~\ref{fig:FEL}), although the
mean structural changes upon ligand switching are rather small.
That is, the secondary and tertiary structure of the protein are quite
similar ($\lesssim\,1\,$\AA\ RMS displacement) in the different
states, and only modest ($\sim\,20\,$\%) shifts of the state's
population are found (Fig.~\ref{fig:FEL}b).
On average, the measurable structural change is therefore only in the
order of 0.2~\AA. In light of this result,
it is remarkable that we can observe such minor structural changes by
transient IR spectroscopy (Fig.~\ref{FigTransientIR}), unpinning the
extraordinary structural sensitivity of the method.

Using isotope labeling to discriminate the dynamics of protein and
ligand, the resulting time-resolved double-difference IR spectra have
revealed complex kinetics of the protein that cover many time-scales
(Fig.\ \ref{FigTransientIR}).  The spectra for
\textit{trans}-to-\textit{cis} and \textit{cis}-to-\textit{trans}
ligand-switching are not mirror-images from each other, and the
\textit{trans}-to-\textit{cis} signals exhibit short-time transients
that are not found for \textit{cis}-to-\textit{trans}. Moreover, the
\textit{cis}-to-\textit{trans} transition does not seem to be finished
within 42~$\mu$s (Fig.\ \ref{FigTransientIR}c).  The overall slower
response of the \textit{cis}-to-\textit{trans} transition reflects the
general observation that enforced leaving of a well-defined (low entropy)
ligand binding structure (here \textit{trans}) occurs faster than starting in a
conformationally disordered (high-entropy) state (here \textit{cis}) and
trying to find stabilizing interactions to end in a more organized
structure.\cite{Wolf2019}

More specifically, the {\em trans}-to-{\em cis} non-equilibrium
simulations reveal that the ligand remains bound with its C-terminus
to the protein binding site between $\beta_2$ and $\alpha_2$ up to
about 1~$\mu$s. In this way, it stabilizes the main bound protein
conformation (state {\bf 1}). At longer times, it starts to move out
from the binding pocket, but remains non-specifically bound to the
protein surface. While diffusion on the surface may continue for long
times after {\em trans}-to-{\em cis} switching, it only little affects
the protein internal structure. Nevertheless, this diffusion will be
the first rate-limiting step after \textit{cis}-to-\textit{trans}
switching, which might be the reason that the ligand does not
completely localize in the binding pocket within $42\,\mu$s.

The existence of well-defined metastable conformational states implies
a time-scale separation between fast intrastate fluctuations and rarely
occurring interstate transitions. This allowed us to construct a Markov
state model (MSM), which illustrates the connectivity and transition
times between the metastable states (Fig.\ \ref{fig:MSM}d). In
particular, the discrete time-scales predicted by the MSM are directly
reflected in the dynamical content calculated for experiments and MD
simulations (Fig.\ \ref{Fig1D}e-h), which both cover time-scales from
$\sim\,$1~ns to 10~$\mu$s. Reflecting different observables
(transition dipole vs.\ C$_\alpha$-distances, respectively), the
weights of the various peaks are different.

While ligand switching was shown to cause a conformational transition
of PDZ2 in terms of the mean structure, at the same time it may also
effect a change of the protein's fluctuations. Comparing the time
evolution of the means of the distances and their RMSD, Figs.\
\ref{fig:MSM}a,b reveal that the two quantities correlate closely, a
behavior that is found for all considered C$_\alpha$-distances (Supplementary Fig.\
\SIdistances). This finding reflects the fact that the
C$_\alpha$-distance distributions pertaining to the individual states
are in most cases well separated (Supplementary Fig.\ \SIstateDistanceDist), such
that a transition between two states affects both mean and
variance. Accounting for an entropic contribution of the
conformational transition, the latter effect is often referred to as
``dynamic allostery''.\cite{Cooper84,Fuentes06,Petit09} The
above findings indicate that allosteric transitions may involve both,
conformational and dynamic changes in the case of the PDZ2
domain.\cite{Nussinov15} The answer to what is the dominant effect
will greatly depend on the system under consideration and on the
applied experimental method. While the overall structural change
($\lesssim\,0.2\,$\AA\ RMS displacement) may be too small to be detected
by structure analysis, NMR relaxation methods can sensitively explore
the structural flexibility of proteins. The IR spectrum of the amide I
band, in contrast, is commonly thought of as a measure of
structure,\cite{barth02} but dephasing due to fast fluctuation might
also affect the IR lineshape.

In conclusion, we have characterized the non-equilibrium allosteric
transition in a joint experimental-theoretical approach. The protein
\textit{per se} was kept unmodified, hence ligand-switching mimics very closely the naturally occurring
allosteric perturbation caused by ligand (un)binding events.
We employed a widely studied model system for this purpose, the PDZ2 domain, which is small enough to allow for a  characterization of the process in atomistic detail by MD simulations, but we believe that the findings are of more general nature. That is, while the ligand-induced allosteric transition originates from a
  population shift between various metastable conformational states,
  the measurable mean structural change of the protein may be tiny and
  therefore difficult to observe \cite{Nussinov15}. Moreover, we
  suggest that the separation between purely dynamically driven
  allostery and allostery upon a conformational change may not be as
  clear-cut as previously thought,
but rather that there may be an interplay
between both that allows proteins to adapt their free energy
  landscape to incoming signals. The photo-switching approach presented here is very versatile, and allows us to shed light on the aspects of ``time'' and ``speed'' in allosteric communication.

\section*{Materials \& Methods}


\subsection{Protein and Peptide Preparation} \label{ProteinPrep}
Expression of the wild type PDZ2 domain from human phosphatase 1E,\cite{kozlov02} isotope labelled ($^{13}$C$^{15}$N) protein variant and synthesis of the photoswitchable peptide ligand  was performed as described earlier.\cite{Zanobini2018,Buchli13} The wild type RA-GEF-2 sequence was modified in order to enable cross-linking the photoswitch, while preserving residues that are important for regulation and binding. That is, amino acids at positions (-1) and (-6) were chosen as anchoring points for the photoswitch and mutated into cysteine residues. Four N-terminal residues (RWAK) were added to the sequence in order to improve the water solubility and facilitate the concentration determination of the construct. Final sequence of the peptide was RWAKSEAKECEQVSCV. The purity of all samples was confirmed by mass spectrometry analysis (Fig.~\SIFigMass). All samples were dialyzed against \mbox{50 mM} borate, \mbox{150 mM} NaCl buffer, \mbox{pH = 8.5}. For transient infrared measurements, samples were lyophilized and resuspended in D$_2$O. Incubation of the samples in D$_2$O overnight at room temperature before the measurements eliminated H/D exchange during experiments. The concentration of the samples was determined via the tyrosine absorption at 280~nm for the protein and 310~nm for the peptide and confirmed by amino-acid analysis.

\subsection{Determining the Binding Affinity}
Isothermal titration calorimetry (ITC) measurements were performed on a MicroCal ITC200 (Malvern, UK). In order to ensure the obtained values for the \textit {cis} and \textit{trans} measurement were mutually  comparable, the experiments were performed using the same stock solution of the peptide and protein for both measurements, and under exactly the same experimental conditions. The experiment was performed in triplicate in order to ensure the reproducibility of the data. The sample cell was loaded with 250~$\mu$l of 80~$\mu$M PDZ2 domain solution and the syringe was loaded with 40~$\mu$l of 800~$\mu$M photoswitchable peptide solution. For the \textit {trans} measurement, the system was kept in the dark for the duration of the experiment, while for the \textit {cis} measurement the syringe was constantly illuminated with a 370~nm cw laser (CrystaLaser, power \mbox{$\approx$ 90 mW}).\cite{Jankovic2019} The results are shown in Fig.~\SIFigITC.

As alternative method to determine the binding affinity, we also used circular dichroism  (CD) spectroscopy as well as fluorescence quenching. Both spectroscopic signals change upon the formation of a protein-ligand complex, hence, when measuring them in dependence of peptide and protein concentration, the binding affinity can be fitted assuming a bimolecular equilibrium. CD measurements were done on Jasco (Easton, MD) model J810 spectropolarimeter in a 0.1~cm quartz cuvette as described previously.\cite{Jankovic2019}. Intrinsic tryptophan fluorescence quenching experiment was done on PelkinElmer spectrofluorimeter as described previously.\cite{Jankovic2019} In either case, the protein concentration was kept constant at 5~$\mu$M, respectively, while the peptide concentrations were varied. Fig.~\SIFigCDFlou~ shows the results for the CD spectroscopy and trypthophan fluorescence quenching, while Table~\SITableKd~ compares the binding affinities obtained from all different methods.

\subsection{Transient IR Spectroscopy}
Transient VIS-pump-IR-probe spectra were recorded using two electronically synchronized Ti:Sapphire laser systems\cite{Bredenbeck2004} running at 2.5 kHz. The wavelength of the pump-laser was tuned as to obtain 380~nm pump pulses (2.1~$\mu$J) for the \textit{trans}-to-\textit{cis} experiment, and 420~nm (1.3~$\mu$J) for the \textit{cis}-to-\textit{trans} experiment, respectively, \textit{via} second harmonic generation in a BBO crystal. The beam diameter of the pump pulse at the sample position was $\approx$180~$\mu$m, employing a pulse duration of $\approx$200~ps (by extracting the light directly after the regenerative amplifier and before the compressor) to minimize the sample degradation during the measurements. Mid-IR probe pulses centered at $\approx$1630~cm$^{-1}$ (pulse duration $\approx$100~fs,  beam diameter on the sample $\approx$150~$\mu$m) were obtained in a optical parametric amplifier (OPA),\cite{ham00b} passed through a spectrograph and detected in a 2$\times$64 MCT array detector with a spectral resolution of $\approx$2~cm$^{-1}$/pixel. Pump-probe spectra were acquired up to the maximum delay value of $\approx$42~$\mu$s  with a time resolution of $\approx$200~ps. Normalisation for noise suppression was performed as described in Ref.~\cite{Feng2017}.

The samples ($\approx$700~$\mu$l) were pumped through a closed flow-cell system purged with N$_2$. The system consisted of a sample cell with two CaF$_2$ windows separated by a 50~$\mu$m Teflon spacer and a reservoir. The flow speed in the sample cell was optimized in order to minimize loss of sample at the largest pump-probe delay time ($\approx$42~$\mu$s) on the one hand, but to have the sample exchanged essentially completely for the subsequent laser shot after 400~$\mu$s on the other hand. The concentrations of the samples were set at 1.25~mM for the peptide and 1.5~mM for the protein. A slight excess of protein was needed to ensure that the peptide was fully saturated with the protein; in order to eliminate the response of  free, photoswitchable peptide. As a reference, FTIR difference spectra have been taken in a Bruker Tensor 27 FTIR spectrometer, using the same sample conditions.

For the experiment with \textit{trans}-to-\textit{cis} switching, we relied on thermal \textit{cis}-to-\textit{trans} back reaction. By comparing its rate with the isomerization probability induced by the 380~nm pump light (determined by pump light power, total sample volume, absorption cross sections,\cite{zhang03}
and isomerization quantum yield\cite{borisenko05}), we estimated that the photo-equilibrium in the total sample volume is 70\%/30\%  \textit{trans}/\textit{cis} during measurement. It furthermore helps that the absorption cross section at 380~nm of the azobenzene moiety in the \textit{trans}-state is $\approx$20 times larger than that of the \textit{cis}-state,\cite{zhang03} which leads us to conclude that $>$97\% of the molecules in the \textit{trans}-to-\textit{cis} experiment undergo the desired isomerisation direction.

For the experiment with \textit{cis}-to-\textit{trans} switching, the sample could be actively switched back by illuminating the reservoir with an excess of light at 370~nm from a cw laser (CrystaLaser, 150~mW).

\subsection{MD Simulations}

All MD simulations of PDZ2 were performed using the GROMACS v2016
software package\cite{Abraham15} and the Amber99*ILDN force
field.\cite{Hornak2006, Best2009, LindorffLarsen2010} Force field parameters
of the azobenzene photoswitch were taken from
Ref.~\cite{Zanobini2018}. Protein-ligand structures were solvated
with ca.\ 8000 TIP3P water molecules\cite{Jorgensen1983} in a
dodecahedron box with a minimal image distance of 7 nm.  16
Na$\textsuperscript{+}$ and 16 Cl$\textsuperscript{-}$ were added to
yield a charge-neutral system with a salt concentration of 0.1 M. All
bonds involving hydrogen atoms were constrained using the LINCS
algorithm,\cite{Hess2008} allowing for a time step of 2 fs. Long-range
electrostatic interactions were computed by the Particle Mesh Ewald
method,\cite{Essmann1995} whereas the short-range electrostatic
interactions were treated explicitly with the Verlet cutoff
scheme. The minimum cutoff distance for electrostatic and van der
Waals interactions was set to 1.4 nm. A temperature of 300 K was
maintained via the Bussi thermostat\cite{Bussi2007} (aka
velocity-rescale algorithm) with a coupling time constant of
$\tau_{T}$ = 0.1 ps. A pressure $P = $1 bar was controlled using the
pressure coupling method of Berendsen\cite{Berendsen1984} with a
coupling time constant of $\tau_{P}$ = 0.1 ps.


The starting structure of the photoswitched ligand bound to PDZ2 was
prepared previously (see Ref.~\cite{Zanobini2018}) based on the
crystal structure (PDB ID 3LNX\cite{Zhang10}). Here, the azobenzene
photoswitch was attached in $\textit{trans}$ conformation to the
ligand at positions (-6) and (-1), which had been mutated to cysteins
as in experiment to provide covalent connection points. Residues
missing at the N-terminus of the ligand were added (see Sec.\
\ref{ProteinPrep}). Following NPT equilibration of the system in
$\textit{trans}$ conformation for 10~ns, 4 statistically independent
(i.e., with different initial velocity distributions) NVT runs of
100~ns each were performed.  For one, we selected 5 randomly chosen
snapshots from the end of these trajectories to perform
$5\!\times \!5\mu$s-long {\bf $\textit{trans}$ equilibrium}
simulations. Moreover, we selected 25 randomly chosen snapshots from
each of the last 50~ns of these four NVT trajectories to perform
\textit{trans}-to-\textit{cis} nonequilibrium simulations, yielding a
total of 100 starting structures which consists mostly of
metastable state 1 (for state definition, see Sec.\ \ref{sec:PCA}).
Employing these initial conditions, \textit{trans}-to-\textit{cis}
photoswitching was performed using a previously developed
potential-energy surface switching approach \cite{Nguyen06}. All 100
{\bf \textit{trans}-to-\textit{cis} nonequilibrium} simulations were
run for $1 \mu$s; 25 of them were extended to a length of
10~$\mu$s.

Upon switching the ligand from {\em trans} to {\em cis} configuration,
PDZ2 undergoes a nonequilibrium time evolution until it relaxes within
a few microseconds (see below) into its {\em cis} equilibrium state,
describing the unbound protein-ligand complex. Performing
$25 \!\times\! 10\, \mu$s-long {\em trans}-to-{\em cis} nonequilibrium
simulations, we took the last $7 \mu$s of each trajectory to estimate
the rather heterogeneous conformational distribution of the {\bf {\em
  cis} equilibrium} state.
To generate initial structures for \textit{cis}-to-\textit{trans}
photoswitching, we took from the 25 \textit{trans}-to-\textit{cis}
trajectories 100 randomly chosen snapshot at a simulation time around
3.0$\mu$s.
Following photoswitching, 100 {\bf \textit{cis}-to-\textit{trans}
  nonequilibrium} trajectories were simulated for a trajectory length
of 1~$\mu$s; 10 simulations were extended to a length of 8~$\mu$s.

Gromacs tools $gmx~angle$ and $gmx~mindist$ were employed to compute
backbone dihedral angles, interresidue $C_{\alpha}$-distances, and the
number of contacts between various segments of PDZ2. Time-dependent
distributions and mean values of these observables were calculated via
an ensemble average over 100 nonequilibrium trajectories.

%
%
\subsection{Dimensionality reduction and clustering} \label{sec:PCA}


To choose suitable internal coordinates that account for the
conformational transitions of the system, \cite{Sittel2018} we
determined 56 $C_{\alpha}$-distances $d_{i,j}$ between residues $i$
and $j$ that are not redundant (such as $d_{i,j}$ and $d_{i,j\pm 1}$)
and whose ensemble average changes significantly
($\langle d_{ij} \rangle \ge 0.5\,$\AA) during the first microsecond
\textit{trans}-to-\textit{cis} nonequilibrium simulations, see Fig.\
\SIdistances. Moreover, we considered all backbone dihedral angles
that show a change of $\gtrsim 10^\circ$ from their initial value
during the \textit{trans}-to-\textit{cis} nonequilibrium simulations.

Since the interresidue
$C_{\alpha}$-distances appear to provide more information, these
coordinate are chosen for the subsequent principal component analysis
(PCA), which was performed on all data.\cite{Sittel2018} For adequate
relative weighting of short and long distances, the data was
normalized. \cite{Ernst2015} Diagonalizing the resulting covariance
matrix, we obtain its eigenvectors (yielding the PCs) and eigenvalues
(reflecting the fluctuations of the PCs). The first two PCs cover 43 \% of
the overall fluctuations, while six PCs yield about 65 \%.
Calculating the free energy profiles pertaining to the PCs, we find
that in particular PC 1--4, 6 and 7 show multistate behavior
reflecting metastable states.

Including these 6 PCs, we performed
robust density-based clustering,\cite{Sittel16} which first computes a
local free energy estimate for every structure in the trajectory by
counting all other structures inside a $6$-dimensional hypersphere of
fixed radius $R$. Normalization of these population counts yields
densities or sampling probabilities $P$, which give the free energy
estimate $\Delta G = -k_\text{B}T \ln P$. Thus, the more structures
are close to the given one, the lower the free energy estimate. By
reordering all structures from low to high free energy, finally the
minima of the free energy landscape can be identified. By iteratively
increasing a threshold energy, all structures with a free energy below
that threshold that are closer than a certain lumping radius will be
assigned to the same cluster, until all clusters meet at their energy
barriers. In this way, all data points are assigned to a cluster as
one branch of the iteratively created tree. For PDZ2, we used a
hypersphere $R = 0.579$ that equaled the lumping radius employed in
the last step.

Figure \SIDBC(top) shows the resulting total number of states
obtained as a function of the minimal populations $P_{min}$ a state
must contain. Here we chose $P_{min}=50~000$, resulting in a
clustering into 12 states. According to visual inspection of the
resulting free energy landscapes (Fig.\ \SIDBC(middle)), these
states separate accurately all density maxima of the system.
Since the 5 lowest populated states cover less than 5 \% of the total
population, we lumped them to main states {\bf 1} to {\bf 7} as
follows: ({\bf 1,\,9})$\rightarrow${\bf 1}, ({\bf
  2,\,10})$\rightarrow${\bf 2}, ({\bf 4,\,12})$\rightarrow${\bf 4},
({\bf 5,\,8,\,11})$\rightarrow${\bf 5}. This is justified due to their
geometric vicinity in the free energy landscape (Fig.\
\SIDBC(middle)), as well as due to their kinetic vicinity in the
transition matrix.
Following the calculation of the time-dependent states populations, in
a last step we lumped states ({\bf 4,\,7})$\rightarrow${\bf 4} and
states ({\bf 5,\,6})$\rightarrow${\bf 5} for the sake of easy
interpretability.

Finally we employed a recently proposed machine learning
approach\cite{Brandt2018} to identify the internal coordinates that
allow to discuss the 5 main states of PDZ2 in a two-dimensional free
energy landscape. On the basis of the decision-tree based program
XGBoost,\cite{Chen2016a} we trained a model that determines the
features of the molecular coordinates that are most important to
discriminate given metastable states. Using a new algorithm that
exploits this feature importance via an iterative exclusion principle,
we identified the essential internal coordinates, that is, the most
important C$_\alpha$-distances of PDZ2. Figure \SIDBC(bottom)
shows that three distances, $d_{20,71}$, $d_{4,55}$ and $d_{27,69}$
suffice to qualitatively distinguish the 5 main states of PDZ2.  The
XGBoost parameters are chosen as in Ref.\ \cite{Brandt2018},
including learning rate $\eta=0.3$, maximum tree depth of 6, 10
training rounds, and 70\% and 30\% of the data used for training and
validation, respectively.

%
%
\subsection{Markov state model} \label{sec:MSM}

On the basis of the above defined 7 metastable states, we constructed
a Markov state model\cite{Bowman2013} of the {\em trans}-to-{\em cis}
transition of PDZ2, using all ($75 \!\times\! 1\,\mu$s and
$25 \!\times\! 10\,\mu$s) {\em trans}-to-{\em cis} nonequilibrium
trajectories.
A general problem with the definition of metastable states is that, due
to the inevitable restriction to a low-dimensional space combined with
insufficient sampling, we often obtain a misclassification of sampled
points in the transition regions, which causes intrastate fluctuations
to be mistaken as interstate transitions. As a simple but effective
remedy, we use dynamical coring which requires that a transition must
a minimum time $\tau_{\rm cor}$ in the new state for the transition to
be counted.\cite{Jain2014,Nagel2019}
A suitable quantity that reflects these spurious crossings is the
probability $W_i(t)$ to stay in state $i$ for duration $t$ (without
considering back transitions). As shown in Fig.\
\SIMSM, without coring we observe a strong initial decay of
$W_i(t)$ for all states, instead of a simple exponential decay we
would expect for Markovian states. Applying coring with increasing
coring times, this initial drop vanishes because fluctuations on
timescales $t \lesssim \tau_\text{cor}$ are removed.
Here we determined
$\tau_{\rm cor} = 1\,$ns as shortest coring time, which removes the
spurious interstate transitions.

Figure \SIMSM shows the resulting implied timescales and
eigenvectors of the model. Using a lag time of 1 ns, we moreover show
the time evolution of the state populations, assuming that we start
completely in a specific state.

\section*{Acknowledgements}
{We thank Rolf Pfister for the synthesis of the peptides and the
Functional Genomics Center Zurich, especially Serge Chesnov and Birgit
Roth, for their help with the mass spectrometry and amino-acid
analysis. We also thank Benjamin Lickert, Daniel Nagel and Georg Diez
for many enlightening discussions concerning the MD data analysis. The
work has been supported by the Swiss National Science Foundation (SNF)
through the NCCR MUST and Grant 200020B\_188694/1, as well as by the
Deutsche Forschungsgemeinschaft through Grant STO 247/10-2. We
acknowledge support by the High Performance and Cloud Computing Group
at the Zentrum f\"ur Datenverarbeitung of the University of T\"ubingen
and the Rechenzentrum of the University of Freiburg, the state of
Baden-W\"urttemberg through bwHPC and the DFG through Grant Nos. INST
37/935-1 FUGG (RV bw16I016) and INST 39/963-1 FUGG (RV bw18A004), the
Black Forest Grid Initiative, and the Freiburg Institute for Advanced
Studies (FRIAS) of the Albert-Ludwigs-University Freiburg.}

\section*{Author contributions}
O.B., C.Z. and A.G. contributed equally to this work. O.B, S.W., G.S., and P.H. designed research, O.B and B.J prepared samples,  all authors provided data and/or analyzed them, O.B, C.Z., S.W., G.S., and P.H contributed to the writing of the paper.

\end{document}